\documentclass[aps,prd,twocolumn,superscriptaddress,showpacs,amsmath,amssymb]{revtex4}
\usepackage{epsf}

\def\be{\begin{equation}}
\def\ee{\end{equation}}
\def\bea{\begin{eqnarray}}
\def\eea{\end{eqnarray}}
\def\d{\mbox{d}}

\def\p{\partial}

\def\pder#1#2{\frac{\partial #1}{\partial #2}}

\let\phi=\varphi
\let\rho=\varrho
\def\bh{black-hole }

\begin{document}

\title{The Aschenbach effect: unexpected topology changes in motion of
  particles and fluids orbiting rapidly rotating Kerr black holes}

\author{Zden\v{e}k Stuchl\'{\i}k}
   \email{Zdenek.Stuchlik@fpf.slu.cz}   
   \affiliation{Institute of Physics, Silesian University at Opava,
                Bezru\v{c}ovo n\'{a}m. 13, CZ-746 01 Opava, Czech Republic}
   \affiliation{NORDITA, Blegdamsvej 17, DK-2100 Copenhagen, Denmark}
\author{Petr Slan\'{y}}
   \email{Petr.Slany@fpf.slu.cz}
   \affiliation{Institute of Physics, Silesian University at Opava,
                Bezru\v{c}ovo n\'{a}m. 13, CZ-746 01 Opava, Czech Republic}
   \affiliation{NORDITA, Blegdamsvej 17, DK-2100 Copenhagen, Denmark}
\author{Gabriel T\"{o}r\"{o}k}
   \email{terek@volny.cz}
   \affiliation{Institute of Physics, Silesian University at Opava,
                Bezru\v{c}ovo n\'{a}m. 13, CZ-746 01 Opava, Czech Republic}
   \affiliation{NORDITA, Blegdamsvej 17, DK-2100 Copenhagen, Denmark}
\author{Marek A. Abramowicz}
   \email{marek@fy.chalmers.se} 
   \affiliation{Institute of Physics, Silesian University at Opava,
                Bezru\v{c}ovo n\'{a}m. 13, CZ-746 01 Opava, Czech Republic}
   \affiliation{NORDITA, Blegdamsvej 17, DK-2100 Copenhagen, Denmark}
   \affiliation{Theoretical Physics, G{\"o}teborg \& Chalmers Universities,
                S-412 96 G{\"o}teborg, Sweden}

\date{\today}

\begin{abstract}
  Newton's theory predicts that
  the velocity ${\cal V}$ of free test particles on circular orbits around a
  spherical gravity center is a decreasing function of the orbital radius $r$,
  $\d{\cal V}/\d r < 0$. Only very recently, \citet{Asch:2004:ASTRA:} has
  shown that, 
  unexpectedly, the same is {\it not} true for particles orbiting black holes:
  for Kerr black holes with the spin parameter $a>0.9953$, the velocity has a
  positive radial gradient for geodesic, stable, circular orbits in a small
  radial range close to the black hole horizon. We show here that the {\em
    Aschenbach effect} occurs also for non-geodesic circular orbits with
  constant specific angular momentum $\ell = \ell_0 = {\rm const}$. In Newton's
  theory 
  it is ${\cal V} = \ell_0/{\cal R}$, with ${\cal R}$ being the cylindrical
  radius. The equivelocity surfaces coincide with the ${\cal R} = {\rm
    const}$ surfaces which, of course, are just co-axial cylinders. It was
  previously known that in the black hole case this simple topology changes
  because one of the ``cylinders'' self-crosses. We show here that the
  Aschenbach effect is connected to a second topology change that for the
  $\ell = {\rm const}$ tori occurs only for very highly spinning black holes,
$a>0.99979$.
\end{abstract}


\maketitle

\bibliographystyle{apsrev}

\section{Introduction}

Aschenbach \cite{Asch:2004:ASTRA:} found a very interesting and rather
surprising fact 
about the circular orbits of free particles around the Kerr black holes with
high spin. Contrary to what is true for Kerr black holes with a small spin,
for orbits around Kerr black holes with $a>0.9953$ the Keplerian orbital 
velocity ${\cal V}^{(\phi)}_{\rm LNRF}$ measured in locally non-rotating
frames (LNRF) is a non-monotonic function of radius.

In this article we show that there is a corresponding change of behaviour
of orbital velocity in the case of non-Keplerian orbits with constant
specific angular momentum \footnote{Here the {\em specific angular momentum}
  means the axial (conserved) angular momentum per total (conserved) energy
  and we use this term for both particles and fluid elements in
  the whole article.}, 
$\ell (r,\theta)=\mbox{const}$: for low spin black 
holes the radial gradient of the orbital velocity, $\p {\cal V}^{(\phi)}_{\rm
  LNRF}/\p r$, changes its sign once, but for a sufficiently rapidly rotating
black holes, it changes the sign three times.

We discuss the geometrical reason for this puzzling behaviour of orbital
velocity in terms of the von Zeipel surfaces, defined as the surfaces of
constant ${\cal R}(r,\theta)\equiv\ell/{\cal V}^{(\phi)}_{\rm LNRF}$. 
In Newton's physics (Euclidean geometry), the von Zeipel surfaces have
topology of 
co-axial cylinders ${\cal R} = r \sin\theta = \mbox{const}$. In the
black 
hole geometry, the topology of the von Zeipel surfaces is remarkably
different. It was known for a long time that for a non-rotating black hole
one of the von Zeipel surfaces self-crosses at the location of the photon
orbit \cite{Abr-Mil-Stu:1993:PHYSR4:}. We found that the {\em Aschenbach
  effect} is due to a second topology  
change, as the another surface with a cusp together with toroidal surfaces
appear. The second change is strictly connected to the first one, but it
occurs only for very rapidly rotating black holes.

In Section \ref{s2}, we summarize basic relations characterizing the
constant specific angular momentum tori. In Section \ref{s3}, the orbital
velocity 
relative to the LNRF is given and its properties are determined. In Section
\ref{s4}, the notion of von Zeipel radius is introduced and properties of the
von Zeipel surfaces are analyzed. In Section
\ref{concl}, we present discussion and some concluding remarks. 

\section{Constant specific angular momentum tori}\label{s2}
In general, stationary and axially symmetric spacetimes with the line element
\be                                               \label{e1}
     \d s^2=g_{tt}\d t^2 + 2g_{t\phi}\d t \d\phi +
         g_{\phi\phi}\d\phi^2 + g_{rr}\d r^2 + g_{\theta\theta}\d\theta^2,
\ee
the stationary and axisymmetric fluid tori with the stress-energy tensor
$T^{\mu\nu}=(\varrho+p)U^{\mu}U^{\nu} + pg^{\mu\nu}$ are characterized by
4-velocity field
\be                                               \label{e2}
     U^{\mu} = (U^{t},0,0,U^{\phi})
\ee
with $U^{t}=U^{t}(r,\theta),\ U^{\phi}=U^{\phi}(r,\theta)$, and by the
distribution of specific angular momentum 
\be                                               \label{e3}
     \ell=-\frac{U_{\phi}}{U_t}.
\ee
The angular velocity of orbiting matter, $\Omega=U^{\phi}/U^t$, is then
related to $\ell$ by the formula
\be                                               \label{e4}
     \Omega=-\frac{\ell g_{tt}+g_{t\phi}}{\ell g_{t\phi}+g_{\phi\phi}}.
\ee

The tori considered here are assumed to have constant
specific angular momentum,
\be                                               \label{e5}
     \ell=\ell (r,\theta)=\mbox{const}.
\ee
Their structure is determined by equipotential surfaces $W=W(r,\theta)$
defined by  
\cite{Abr-Jar-Sik:1978:ASTRA:, %
       Koz-Jar-Abr:1978:ASTRA:}
\be                                               \label{e6}
     W-W_{\rm in}=\ln\frac{U_{t}}{(U_{t})_{\rm in}}
\ee
with 
\be                                               \label{e7}
     (U_t)^2=\frac{g_{t\phi}^2 - g_{tt}g_{\phi\phi}}{g_{tt}\ell^2 +
  2g_{t\phi}\ell + g_{\phi\phi}};
\ee
the subscript ``in'' refers to the inner edge of the torus. 

The metric components of the Kerr spacetime (with $a>0$) in the
Boyer-Lindquist coordinates are:  
\bea 
     g_{tt} & = & -\frac{\Delta - a^2 \sin^2 \theta}{\Sigma},
                                                  \label{e8}\\ 
     g_{t\phi} & = & -\frac{2ar\sin^2 \theta}{\Sigma}, 
                                                  \label{e9}\\
     g_{\phi\phi} & = & \frac{A\sin^2 \theta}{\Sigma}, 
                                                  \label{e10} 
\eea 
where
\bea 
     \Delta & = & r^2-2r+a^2,                         \label{e11}\\
     \Sigma & = & r^2+a^2 \cos^2 \theta,              \label{e12}\\ 
     A & = & (r^2+a^2)^2-\Delta a^2 \sin^2 \theta.     \label{e13}
\eea 
We make our formulae dimensionless by using the standard $c=G=M=1$ 
convention. The relation (\ref{e4}) for the angular velocity of matter
orbiting the black hole acquires the form 
\be                                               \label{e13.1}
     \Omega = \Omega(r,\theta;a,\ell) = \frac{(\Delta-a^2 \sin^2 \theta)\ell +
       2ar\sin^2 \theta}{(A-2\ell ar) \sin^2 \theta} 
\ee
and the potential W, defined in Eq. (\ref{e6}), has the explicit form
\bea                                                \label{e13.2}
  \lefteqn{W = W(r,\theta;a,\ell)=} \nonumber\\
  & & \frac{1}{2}\ln\frac{\Sigma\Delta\sin^2\theta}{(r^2+a^2-a\ell)^2 
      \sin^2\theta - \Delta(\ell-a\sin^2\theta)^2}. 
\eea 

\section{The orbital velocity in LNRF}\label{s3} 
The locally non-rotating frames are given by the tetrad of 1-forms
\cite{Bar-Pre-Teu:1972:ASTRJ2:, %
              Mis-Tho-Whe:1973:Gra:}  
\bea
     \mathbf{e}^{(t)} & = & \left(\frac{\Sigma\Delta}{A}\right)^{1/2} 
     \mathbf{d}t,                                 \label{e14}\\
     \mathbf{e}^{(r)} & = & \left(\frac{\Sigma}{\Delta}\right)^{1/2} 
     \mathbf{d}r,                                 \label{e15}\\
     \mathbf{e}^{(\theta)} & = & \Sigma^{1/2}\mathbf{d}\theta, 
                                                  \label{e16}\\
     \mathbf{e}^{(\phi)} & = & \left(\frac{A}{\Sigma}\right)^{1/2}\sin\theta 
     (\mathbf{d}\phi-\omega\mathbf{d}t)
                                                  \label{e17} 
\eea
where the angular velocity of LNRF, $\omega = -g_{t\phi}/g_{\phi\phi}$, is 
given by the relation
\be                                               \label{e18} 
     \omega = \frac{2ar}{A}.
\ee   
The azimuthal component of 3-velocity in LNRF reads
\be                                               \label{e19} 
     {\cal V}^{(\phi)}_{\rm LNRF}=\frac{U^{\mu} 
       \mathrm{e}^{(\phi)}_{\mu}}{U^{\nu} \mathrm{e}^{(t)}_{\nu}} = 
     \frac{A\sin\theta}{\Sigma\sqrt{\Delta}} (\Omega-\omega).   
\ee 
Substituting for the angular velocities $\Omega$ and $\omega$ from the
relations (\ref{e13.1}) and (\ref{e18}), respectively, we arrive at the formula
\be                                               \label{e20} 
     {\cal V}^{(\phi)}_{\rm LNRF}=\frac{A(\Delta - a^2 \sin^2 \theta) + 4a^2
      r^2 \sin^2 \theta}{\Sigma\sqrt{\Delta} (A-2a\ell r)\sin\theta}\ell. 
\ee

\begin{figure}
\centering 
\epsfxsize=.8 \hsize                                 
\epsfbox{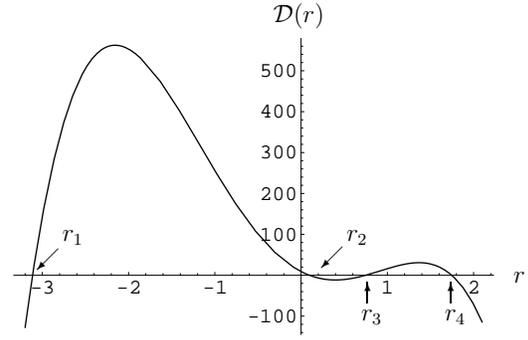} 
\caption{Reality condition for the existence of local extrema of the
  function  
  $\ell_{\rm ex}(r;a)$. The extrema are allowed, if ${\cal D}(r)>0$. Clearly,
  the physically relevant extrema, located above the outer horizon, can exist 
  in the interval $r\in (1,r_4)$.}
\label{f1} 
\end{figure} 

\begin{figure}
\epsfxsize=.8 \hsize                                 
\epsfbox{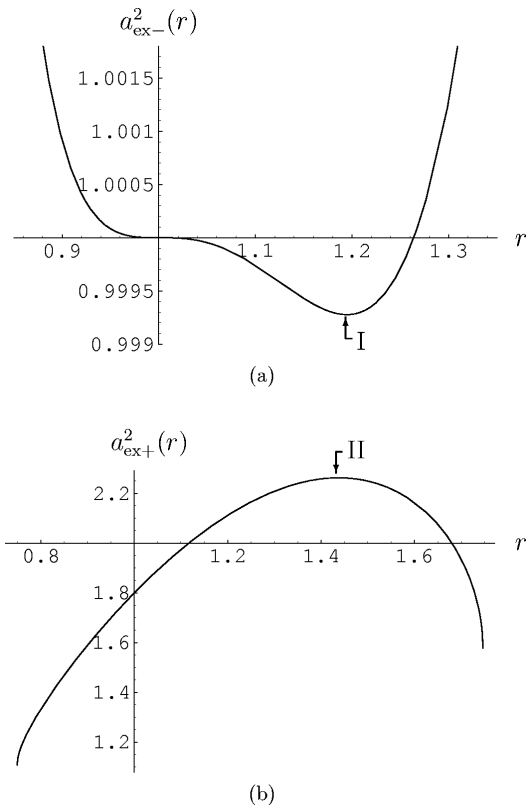} 
\caption{Loci of local extrema of the function $\ell_{\rm 
  ex}(r;a)$. They are  
  determined by the functions $a^2_{\rm ex\pm}(r)$. (a) The function $a^2_{\rm
  ex-}(r)$ is relevant for both black holes and naked singularities; its local 
  minimum is denoted I. (b) The function $a^2_{\rm ex+}(r)$ is relevant for
  naked singularities only; its local maximum is denoted II.} 
\label{f2}
\end{figure} 

In the equatorial plane, $\theta=\pi/2$, equation (\ref{e20}) reduces to  
\be                                               \label{e21}
     {\cal V}^{(\phi)}(r,\theta=\pi/2;a,\ell) = 
     \frac{r\sqrt{\Delta}}{r(r^2+a^2)-2a(\ell-a)}\ell.  
\ee 
The velocity vanishes at infinity, $r\to\infty$, and at the horizon, $r\to
r_{+}=1+\sqrt{1-a^2}$. Thus, a change of the sign of the radial gradient of 
velocity occurs for {\em any} pair $a$, $\ell$. This is not true for  
Keplerian orbits. To see this, one may use the formula for Keplerian angular
velocity \cite{Bar-Pre-Teu:1972:ASTRJ2:}  
\be                                               \label{e21.1}
     \Omega=\Omega_{\rm K}(r;a)=\frac{1}{(r^{3/2}+a)} 
\ee
together with equation (\ref{e19}) to get, in the equatorial 
plane,
\be                                               \label{e22} 
     {\cal V}^{(\phi)}_{\rm K}(r;a)=\frac{(r^2+a^2)^2 - a^2\Delta -
       2ar(r^{3/2}+a)}{r^2(r^{3/2}+a)\sqrt{\Delta}}  
\ee 
Obviously, the velocity formally diverges at $r=r_{+}$. The Keplerian specific
angular momentum is given by  
\be                                               \label{e22.1}
     \ell_{\rm K}(r;a)=\frac{r^2-2ar^{1/2}+a^2}{r^{3/2}-2r^{1/2}+a}. 
\ee
The minimum of $\ell_{\rm K}(r;a)$ corresponds to the marginally stable
circular geodesic at $r=r_{\rm ms}$, which is the innermost possible
circular, stable geodesic. The innermost possible circular fluid orbit in the
constant specific angular momentum tori is given by the condition $\ell =
\ell_{\rm K}(r_{\rm in})$. It is known that $r_{\rm mb} < r_{\rm in} < r_{\rm
  ms}$. Here $r_{\rm mb}$ denotes the radius of the 
circular marginally bound geodesic.

The radial gradient of the equatorial orbital velocity of tori reads
\begin{widetext}
\be                                              \label{e23}
  \pder{{\cal V}^{(\phi)}}{r} = \frac{[\Delta+(r-1)r][r(r^2+a^2)-2a(\ell-a)] -
       r(3r^2+a^2)\Delta}{[r(r^2+a^2)-2a(\ell-a)]^2 \sqrt{\Delta}}\ell,
\ee
\end{widetext}
and it changes its orientation at radii determined 
by the condition
\be                                               \label{e24}
     \ell=\ell_{\rm ex}(r;a) \equiv a +
     \frac{r^2[(r^2+a^2)(r-1)-2r\Delta]}{2a[\Delta+r(r-1)]}.
\ee

We have to discuss properties of $\ell_{\rm ex}(r;a)$ above the event horizon
$r_{+}$ taking into account the limits on the inner boundary of the tori,
$\ell\in(l_{\rm ms},l_{\rm mb})$ where $\ell_{\rm ms}\ (\ell_{\rm mb})$
denotes specific angular momentum of the marginally stable (marginally bound)
circular geodesic. The local extrema of $\ell_{\rm ex}(r;a)$ are given by the
relation 
\be                                             \label{e25}
     a^2=a^2_{\rm ex\pm}(r)\equiv r\frac{3+18r-7r^2\pm\sqrt{{\cal
           D}(r)}}{2(3r+2)},  
\ee
with
\bea                                               \label{e26}
  \lefteqn{{\cal D}(r) = 9-108r+150r^2-12r^3-23r^4 =} \nonumber\\
  & & = -23(r-r_1)(r-r_2)(r-r_3)(r-r_4),
\eea
where
\bea
     r_1 &\doteq & -3.11363, \nonumber\\
     r_2 &\doteq & 0.09602,  \nonumber\\
     r_3 &\doteq & 0.74939,  \nonumber\\
     r_4 &\doteq & 1.74648.  \nonumber
\eea
The situation is illustrated in Fig.~\ref{f1} which implies that only the
interval $r\in(r_3,r_4)$ is relevant for the region outside of the \bh event
horizon. Behaviour of the functions $a^2_{\rm ex\pm}(r)$ is given in
Fig.~\ref{f2}. Clearly, only $a^2_{\rm ex-}(r)$ is relevant for black
holes. The minimum of $a^2_{\rm ex-}(r)$, denoted I, is located at radius
$r_{\rm min}\doteq 1.19466$ and the critical value of the rotational parameter
is 
\be                                              \label{e27}
     a_{\rm c(bh)}\doteq\sqrt{0.99928}\doteq 0.99964. 
\ee
Note that both the functions $a^2_{\rm ex\pm}(r)$ are relevant for Kerr naked
singularities. The maximum of $a^2_{\rm ex+}(r)$, denoted II, is located
at radius $r_{\rm max}\doteq 1.43787$ and the critical value of the rotational
parameter is 
\be                                             \label{e28}
     a_{\rm c(ns)}\doteq\sqrt{2.26289}\doteq 1.50429.
\ee
Therefore, the possibility to have three changes of the sign of $\p {\cal
  V}^{(\phi)}/\p r$ in constant specific angular momentum tori is limited from
bellow for black holes, and from above for naked singularities. Here we
restrict our attention to the Kerr black holes.

\begin{figure}
\centering
\epsfxsize=.99 \hsize                                 
\epsfbox{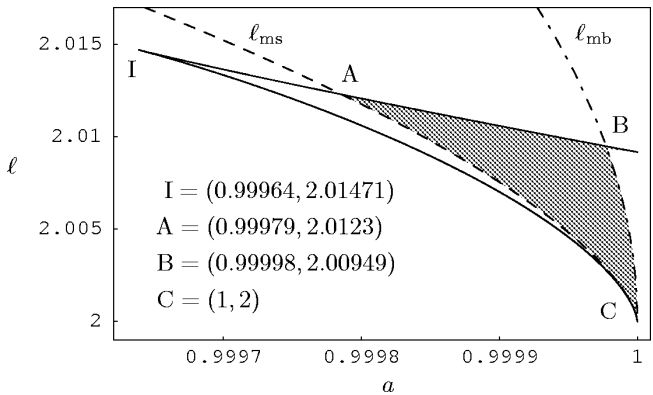}
\caption{Kerr spacetimes with the change of sign of
  the gradient of LNRF velocity. In the $\ell$--$a$ plane, the functions
  $\ell_{\rm ex(max)}(a)$ (upper solid curve), $\ell_{\rm ex(min)}(a)$ (lower
  solid curve), $\ell_{\rm ms}(a)$ (dashed curve) and $\ell_{\rm mb}(a)$
  (dashed-dotted curve) are given. For pairs of $(a,\ell)$ from the shaded
  region, the gradient of the orbital velocity of tori changes its sign
  twice inside the tori. Between the points A, B, the
  $\ell_{\rm ex(max)}(a)$ curve determines 
  an inflex point of ${\cal V}^{(\phi)}(r;a,\ell)$. The inflex points
  determined by the curve $\ell_{\rm ex(min)}(a)$ are irrelevant being outside
  of the definition region for constant specific angular momentum tori,
  $\ell\in (\ell_{\rm ms}(a),\ell_{\rm mb}(a))$. The point I corresponds to
  the inflex point of $\ell_{\rm ex}(r;a)$; cf. Fig.~\ref{f2}a.}
\label{f3}
\end{figure}

Now, we have to compare the local extrema of the function $\ell_{\rm ex}(r;a)$,
determined by the condition (\ref{e25}), with the functions characterizing the
marginally stable, $\ell_{\rm ms}(a)$, and the marginally bound, $\ell_{\rm
  mb}(a)$, circular geodesics as these determine the limits of allowance of
stationary toroidal structures in the Kerr spacetimes
\cite{Abr-Jar-Sik:1978:ASTRA:}. For each given value of $a$, location of both
the marginally stable and the marginally bound circular geodesics 
is uniquely given by the functions $r_{\rm ms}=r_{\rm ms}(a),\ r_{\rm
  mb}=r_{\rm mb}(a)$ \cite{Bar-Pre-Teu:1972:ASTRJ2:}, and
$\ell_{\rm 
  ms}(a),\ \ell_{\rm mb}(a)$ can then be determined using the formulae for
$\ell_{\rm K}(r;a)$ and $r_{\rm ms}(a),\ r_{\rm mb}(a)$, respectively. In
Fig.~\ref{f3}, behaviour of the local extrema $\ell_{\rm ex(min)}(a),\
\ell_{\rm ex(max)}(a)$ and the functions $\ell_{\rm ms}(a),\ \ell_{\rm mb}(a)$
is illustrated. It is clear immediately that the sign's change of $\p {\cal
  V}^{(\phi)}/\p r$ is relevant only for tori orbiting the Kerr black
holes with the rotational parameter
\be                                             \label{e29}
     a>a_{\rm c(tori)}\doteq 0.99979,
\ee
which is much higher than the critical value $a_{\rm c(K)}\doteq 0.9953$
determined by Aschenbach for Keplerian discs
\cite{Asch:2004:ASTRA:}. For 
$a>a_{\rm c(tori)}$ the relevance of $\ell_{\rm ex}(r;a)$ is limited from
bellow by $\ell_{\rm ms}(a)$. There is another critical value of the
rotational parameter, $a=a_{\rm c(mb)}\doteq 0.99998$, where $\ell_{\rm
  mb}(a)=\ell_{\rm ex(max)}(a)$; for $a>a_{\rm c(mb)}$ the relevance of
$\ell_{\rm ex}(r;a)$ is limited from above by $\ell_{\rm mb}(a)$.

The character of the region, where $\p {\cal V}^{(\phi)}/\p r$ changes sign,
can be properly illustrated by considering the functions $\ell_{\rm ex}(r;a)$
and $\ell_{\rm K}(r;a)$ simultaneously. First, we show that there is no common
point of those functions in \bh spacetimes with $a<1$. Indeed, the condition
$\ell_{\rm ex}(r;a)=\ell_{\rm K}(r;a)$ implies an equation quartic in $a$,
which has four solutions
\bea
     a & = & a_1(r) \equiv r\sqrt{r},                        \label{e30} \\
     a & = & a_2(r) \equiv -\sqrt{r(2-r)},                   \label{e31} \\
     a & = & a_{\rm h}(r) \equiv \sqrt{r(2-r)},              \label{e32} \\
     a & = & a_{\rm ph+}(r) \equiv \frac{\sqrt{r}}{2}(3-r).  \label{e33}  
\eea 
The solution $a_1(r)>1$ at $r>1$, i.e., it corresponds to naked singularities
at $r>1$, the solution $a_2(r)$ is negative everywhere, the solution
$a_3=a_{\rm h}(r)$ determines radius of the event horizon, while the solution
$a_4=a_{\rm ph+}(r)$ determines radius of the corotating photon circular
geodesic. 
None of the solutions is relevant for the stationary tori. We can conclude
that above the photon circular orbit there is always $\ell_{\rm
  K}(r;a)>\ell_{\rm 
  ex}(r;a)$; therefore, the innermost local maximum of ${\cal
  V}^{(\phi)}(r;a)$ for $a>a_{\rm c(bh)}$, and the only local maximum of ${\cal
  V}^{(\phi)}(r;a)$ for $a<a_{\rm c(bh)}$, is always physically
irrelevant in constant specific angular momentum tori.    

\begin{figure*}
\centering
\epsfxsize=.735 \hsize                                 
\epsfbox{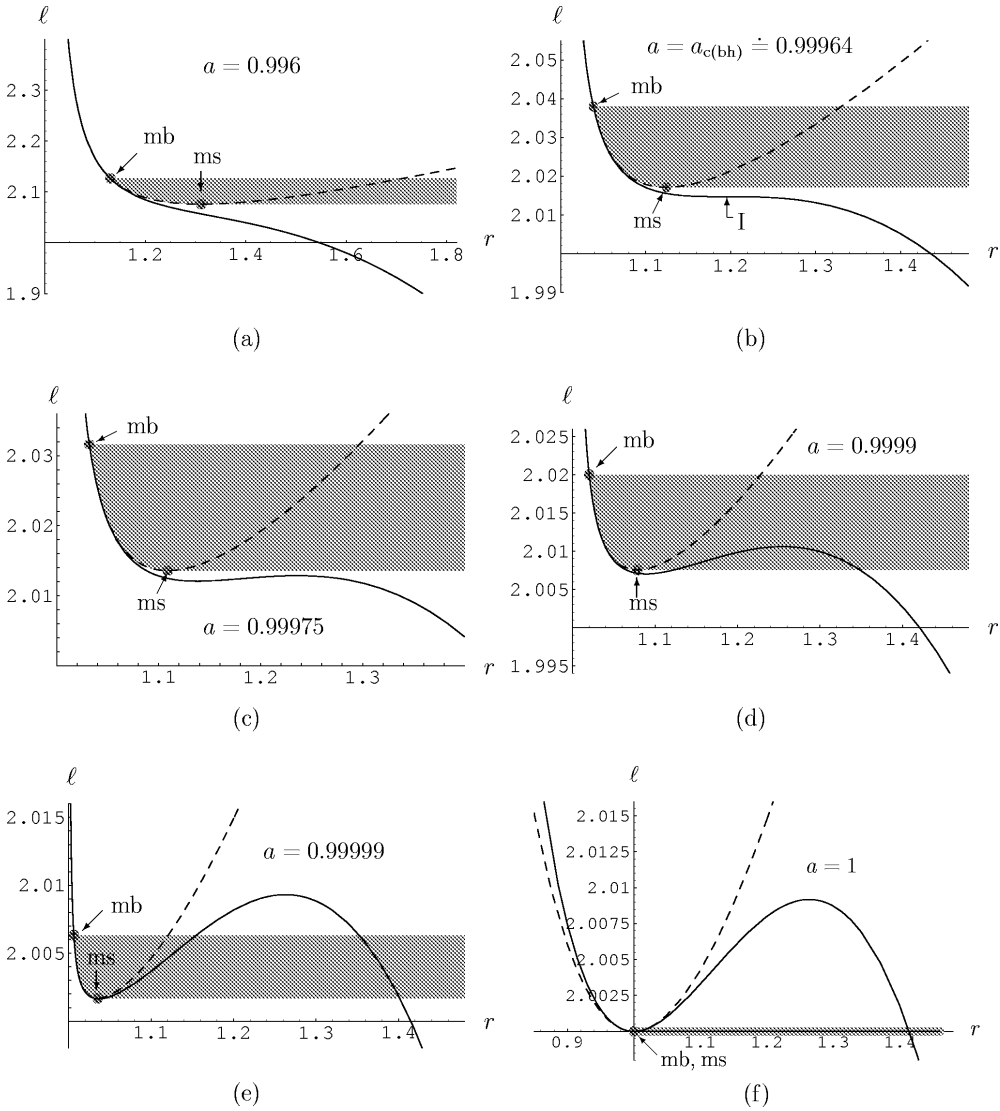}
\caption{Classification of the Kerr \bh spacetimes according to the
  properties of 
  the functions $\ell_{\rm ex}(r;a)$ (solid curves) and $\ell_{\rm K}(r;a)$
  (dashed curves). The functions are plotted for six cases corresponding to
  the classification. The constant specific angular momentum tori can exist in
  the shaded 
  region only along $\ell={\rm const}$ lines. Their inner edge (center) is
  determined by the decreasing (increasing) part of $\ell_{\rm K}(r;a)$. The
  local extrema of the orbital velocity relative to LNRF relevant for tori are
  given by the 
  intersections of $\ell={\rm const}$ line with the curve of $\ell_{\rm
  ex}(r;a)$ in the shaded region. Notice that the region corresponding to the
  allowed values of $\ell$ for the discs is narrowing with $a\to 1$, it is
  degenerated into the $\ell=2$ line for $a=1$ as $\ell_{\rm ms}=\ell_{\rm
  mb}=2$ in this case. In the case (e), the gradient $\p {\cal V}^{(\phi)}/\p
  r$ changes sign for all values of $\ell\in (\ell_{\rm ms},\ell_{\rm
  mb})$ allowed for the tori, while in the case (d), it is allowed for a
  region restricted from above by the value $\ell_{\rm ex(max)}(a)$. In the
  cases (a)--(c), the change of sign of $\p {\cal V}^{(\phi)}/\p r$ cannot
  occur in the disc. It is directly seen from cases (d)--(f) that the gradient
  $\p {\cal V}^{(\phi)}/\p r$ changes the sign closely above the center of the
  disc.}  
\label{f4}
\end{figure*}

For \bh spacetimes, behaviour of the functions $\ell_{\rm ex}(r;a)$  and
$\ell_{\rm K}(r;a)$ can then be classified into six classes which are
illustrated in Fig.~\ref{f4}:
\begin{enumerate}
\item $0<a<a_{\rm c(bh)}$:
No extrema of $\ell_{\rm ex}(r;a)$ (Fig.~\ref{f4}a).
\item $a=a_{\rm c(bh)}$:
An inflex point of $\ell_{\rm ex}(r;a)$ (Fig.~\ref{f4}b).
\item $a_{\rm c(bh)}<a<a_{\rm c(tori)}$:
Two local extrema of $\ell_{\rm ex}(r;a)$ present, but out of the region
allowing the existence of tori (Fig.~\ref{f4}c). 
\item $a_{\rm c(tori)}<a<a_{\rm c(mb)}$:
Two local extrema of $\ell_{\rm ex}(r;a)$ allowed in the region of $\ell\in
(\ell_{\rm ms},\ell_{\rm ex(max)})$ (Fig.~\ref{f4}d).
\item $a_{\rm c(mb)}<a<1$:
Two local extrema of $\ell_{\rm ex}(r;a)$ allowed in the region $\ell\in
(\ell_{\rm ms},\ell_{\rm mb})$ (Fig.~\ref{f4}e).
\item $a=1$:
The minimum of $\ell_{\rm ex}(r;a)$ coincides with the marginally bound
geodesic with $\ell_{\rm mb}=2$ at $r_{\rm mb}=1$. The curves $\ell_{\rm
  ex}(r;a=1)$ and $\ell_{\rm K}(r;a=1)$ intersect at $r=1$ (Fig.~\ref{f4}f). 
\end{enumerate}
Clearly, three changes of sign of $\p {\cal V}^{(\phi)}/\p r$ can occur for
Kerr black holes with the rotational parameter $a>a_{\rm c(bh)}\doteq
0.99964$. However, the effect is relevant for constant specific angular
momentum tori  
only if $a>a_{\rm c(tori)}\doteq 0.99979$. The interval of corresponding
values of the specific angular momentum 
$\ell\in (\ell_{\rm ms}(a),\ell_{\rm ex(max)}(a))$ grows with $a$ growing up
to the critical value of $a_{\rm c(mb)}\doteq 0.99998$. For $a>a_{\rm c(mb)}$,
the interval of relevant values of $\ell\in (\ell_{\rm ms}(a),\ell_{\rm
  mb}(a))$ is narrowing with growing of the rotational parameter up to $a=1$,
which corresponds to a singular case where $\ell_{\rm ms}(a=1)=\ell_{\rm
  mb}(a=1)=2$. Notice that the situation becomes to be singular only in terms
of the specific angular momentum; it is shown
\cite{Bar-Pre-Teu:1972:ASTRJ2:}  
that for $a=1$ both the total energy ${\cal E}$ and the axial angular momentum
${\cal L}$ differ at $r_{\rm ms}$ and $r_{\rm mb}$, respectively, but their
combination, $\ell\equiv {\cal L/E}$, giving the specific angular momentum,
coincides at these radii.  

\begin{figure*}
\epsfxsize=.79 \hsize
\epsfbox{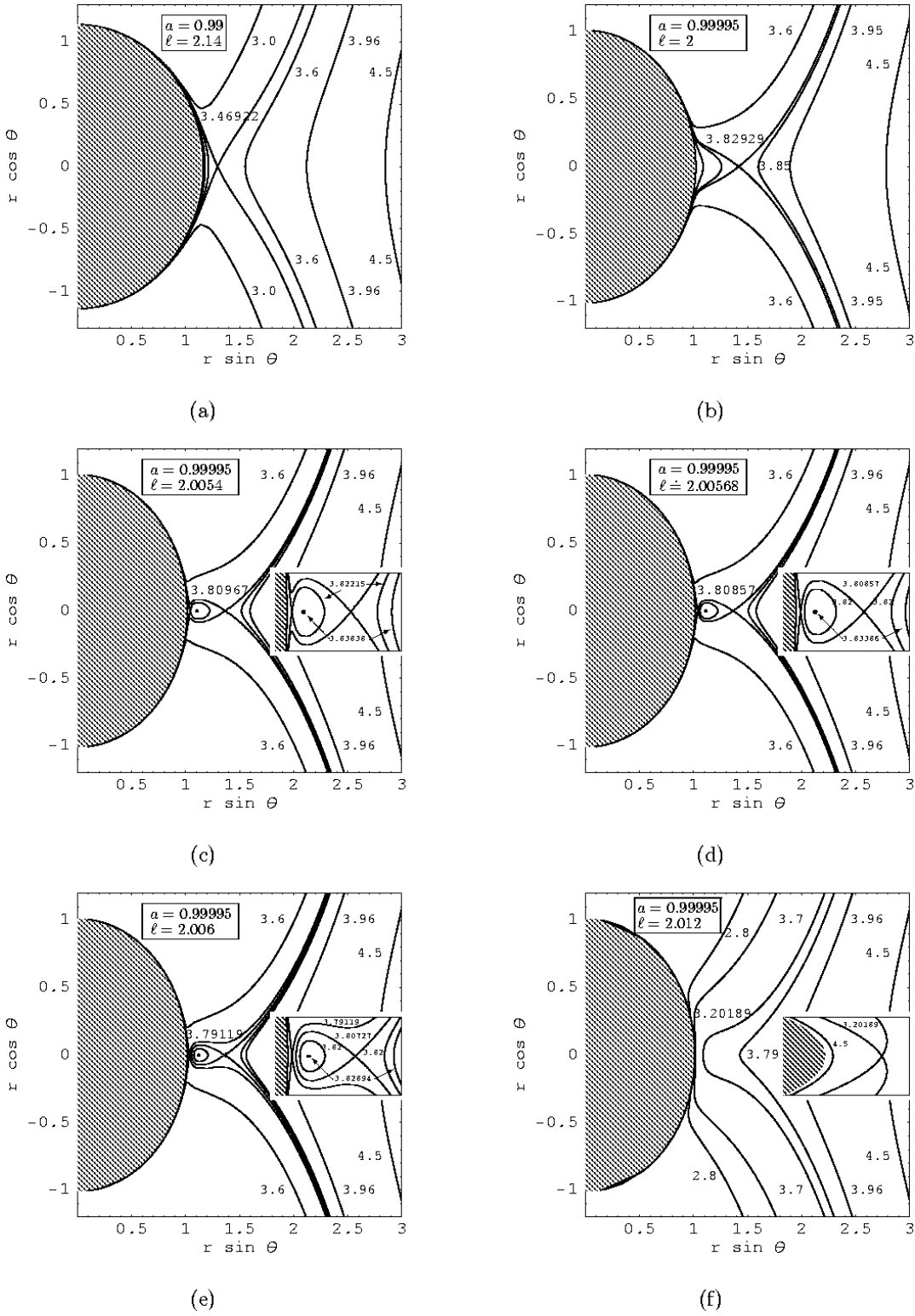}
\caption{Von Zeipel surfaces (meridional sections). (a) For $a<a_{\rm c(bh)}$
  and any $\ell$, only one surface with a cusp in the equatorial plane and no
  closed (toroidal) surfaces exist. The cusp is, however, located outside the
  toroidal equilibrium configurations of perfect fluid. (b)--(f) For
  $a>a_{\rm c(bh)}$ and $\ell$ appropriately chosen, two surfaces with a cusp
  ((c), (e)), or one surface with both the cusps (d), together with closed
  (toroidal) surfaces, exist. Moreover, if $a>a_{\rm c(tori)}$, both the outer
  cusp and the central ring of closed surfaces are located inside the toroidal
  equilibrium configurations corresponding to constant specific angular
  momentum discs  
  (cases (c)--(e)). If $\ell$ is sufficiently low/high ((b)/(f)), there is
  only one surface with a cusp outside the configuration. Shaded region
  corresponds to the black hole.}
\label{f5}
\end{figure*}

\section{Von Zeipel surfaces}\label{s4}

Till now, our study was focused on the properties of LNRF velocity profile in
the equatorial plane of the constant specific angular momentum tori. However,
it is useful to 
obtain global characteristics of the phenomenon that is shown to be manifested
in the equatorial plane as the existence of a small region with positive
gradient of the LNRF velocity.

It is well known that rotational properties of perfect fluid equilibrium
configurations in 
strong gravity are well represented by the radius of gyration $\widetilde{\cal
  R}$, 
introduced in the case of spherically symmetric Schwarzschild spacetimes in
\cite{Abr-Mil-Stu:1993:PHYSR4:}, as the direction of increase of
$\widetilde{\cal R}$ defines a local outward direction of the dynamical
effects of 
rotation of the fluid. In the stationary and axisymmetric spacetimes, the
radius of gyration was defined by the relation \cite{Abr-Nur-Wex:1995:CLAQG:}
\be                                                      \label{e34}
     \widetilde{\cal R} \equiv
     \left(\frac{\tilde{\ell}}{\widetilde{\Omega}}\right)^{1/2}, 
\ee
where $\widetilde{\Omega}=\Omega-\omega$ is the angular velocity relative to
the LNRF. However, $\tilde{\ell}\equiv {\cal L}/\widetilde{\cal E}$ is not the
specific angular momentum $\ell\equiv {\cal L/E}$ with ${\cal L} =
p_{\mu}\xi^{\mu}_{(\phi)},\ {\cal E} = -p_{\mu}\xi^{\mu}_{(t)}$ being the
4-momentum projections on the Killing vector fields $\xi^{\mu}_{(\phi)} =
\delta^{\mu}_{\phi}$ and $\xi^{\mu}_{(t)} = \delta^{\mu}_{t}$, but
$\widetilde{\cal E} = -p_{\mu}\tilde{\eta}^{\mu}$, where $\tilde{\eta}^{\mu} =
\xi^{\mu}_{(t)}+\omega\xi^{\mu}_{(\phi)}$ is not a Killing vector field, i.e.,
  $\widetilde{\cal E}$ is related to the LNRF and it is not a constant of
motion. Important consequence of such a 
definition is given by the relation between $\widetilde{\cal R}$ and ${\cal
  V}^{(\phi)}_{\rm LNRF}$: 
\be                                                      \label{e35} 
     {\cal V}^{(\phi)}_{\rm LNRF}=\widetilde{\Omega}\widetilde{\cal R}.
\ee

Here, we shall use another physically reasonable way of defining a global
quantity characterizing rotating fluid configurations by using directly the
LNRF orbital velocity. We define, so-called, von Zeipel radius by the relation 
\be                                                      \label{e36}
     {\cal R}\equiv\frac{\ell}{{\cal V}^{(\phi)}_{\rm LNRF}}
\ee
which generalizes the Schwarzschildian definition of gyration radius. In static
spacetimes, the von Zeipel radius (\ref{e36}) coincides with the radius of
gyration defined by the relation (\ref{e34}), however, in stationary,
axisymmetric spacetimes, relation between the both radii has the form
\be                                                      \label{e37}
     {\cal R}=(1-\omega\ell)\widetilde{\cal R}.
\ee

In the case of tori with $\ell(r,\theta)=\mbox{const}$, the
von Zeipel surfaces, i.e., the surfaces of ${\cal R}(r,\theta;a,\ell) =
\mbox{const}$, coincide with the equivelocity surfaces ${\cal V}^{(\phi)}_{\rm
  LNRF}(r,\theta;a,\ell)=\mbox{const}$. For the tori in the Kerr spacetimes,
there is
\be                                                      \label{e38}
     {\cal R}(r,\theta;a,\ell) = \frac{\Sigma\sqrt{\Delta}(A-2a\ell
       r)\sin\theta}{A(\Delta-a^2\sin^2\theta)+4a^2r^2\sin^2\theta}.
\ee
Topology of the von Zeipel surfaces can be directly determined by the
behaviour of the von Zeipel radius (\ref{e38}) in the equatorial plane
\be                                                      \label{e39}
     {\cal R}(r,\theta=\pi/2;a,\ell) =
     \frac{r(r^2+a^2)-2a(\ell-a)}{r\sqrt{\Delta}}.
\ee
The local minima of the function (\ref{e39}) determine loci of the cusps of
the von Zeipel surfaces, while its local maximum (if it exists) determines a
circle around which closed toroidally shaped von Zeipel surfaces are
concentrated (Fig.~\ref{f5}). Notice that the minima (maximum) of
${\cal R}(r,\theta=\pi/2;a,\ell)$ correspond(s) to the maxima (minimum) of
${\cal V}^{(\phi)}_{\rm LNRF}(r,\theta=\pi/2;a,\ell)$, therefore, the inner
cusp is always physically irrelevant being located outside of the toroidal
configuration of perfect fluid, cf. Fig.~\ref{f4}. 

\section{Discussion and conclusions}\label{concl}

\begin{figure*}
\epsfxsize=.79 \hsize
\epsfbox{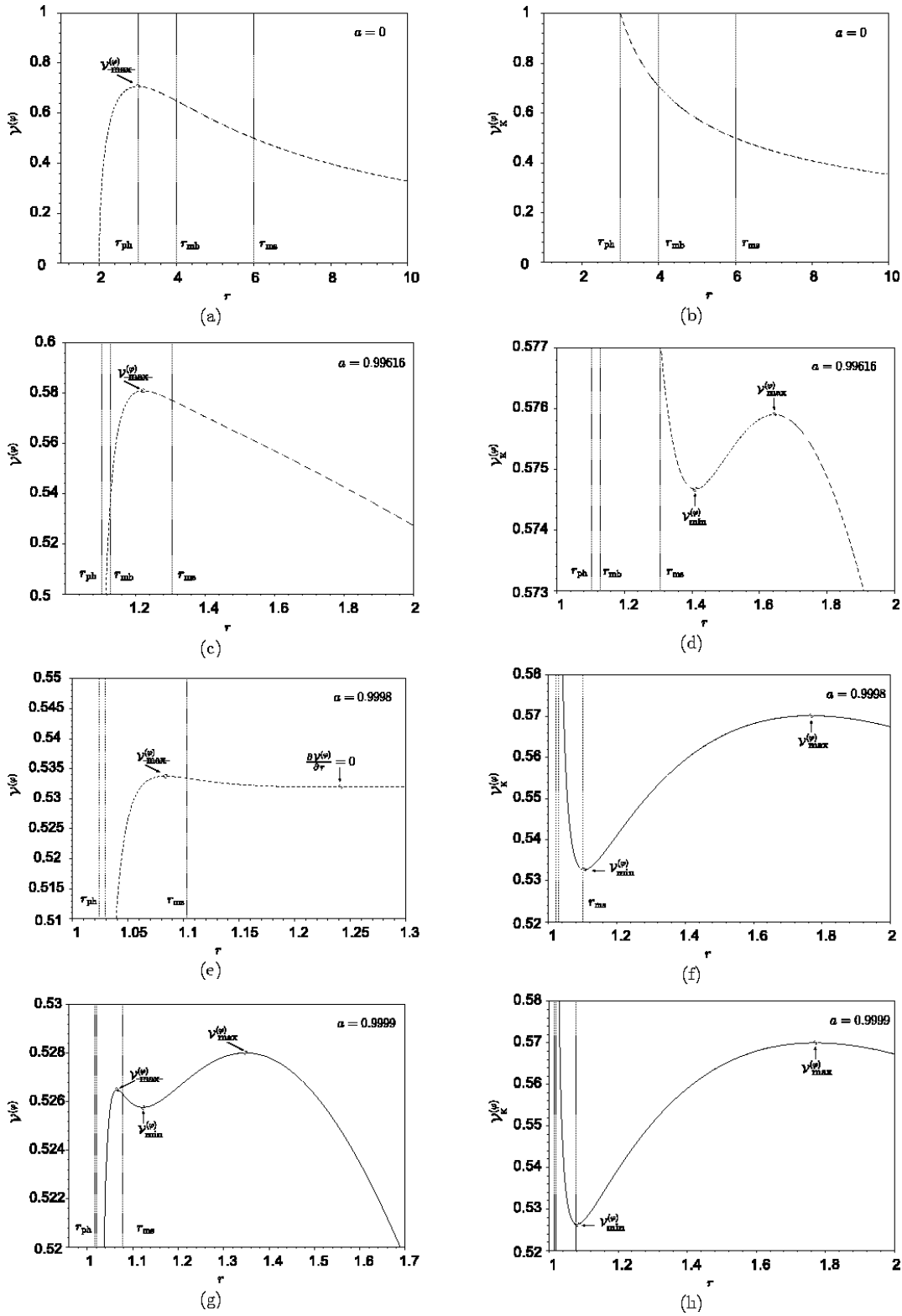}
\caption{Profiles of the equatorial orbital velocity of constant specific
  angular momentum tori  
  in LNRF in terms of the radial Boyer-Lindquist coordinate. The profiles are
  given for typical values of $a$ corresponding to the 
  classification of the Kerr \bh spacetimes. For comparison, the profiles are
  given 
  for the orbital velocity of Keplerian discs in Kerr spacetimes with the same
  rotational parameter $a$. For tori, values of $\ell={\rm const}$ are
  appropriately chosen; commonly, $\ell=\ell_{\rm ms}$ is used giving the
  maximal value of the velocity difference in between the local extrema, and
  representing the limiting case of constant specific angular momentum tori.}
\label{f6}
\end{figure*}

It is useful to discuss both the qualitative and quantitative aspects of the
phenomenon of the positive gradient of LNRF orbital velocity.
In the Kerr spacetimes with $a>a_{\rm c(tori)}$, changes of sign of the
gradient of ${\cal V}^{(\phi)}(r;a,\ell)$ must occur closely above the center
of relevant toroidal discs, at radii corresponding to stable circular
geodesics of the spacetime (cf. Fig.~\ref{f4}).

\begin{figure*}
\epsfxsize=.99 \hsize
\epsfbox{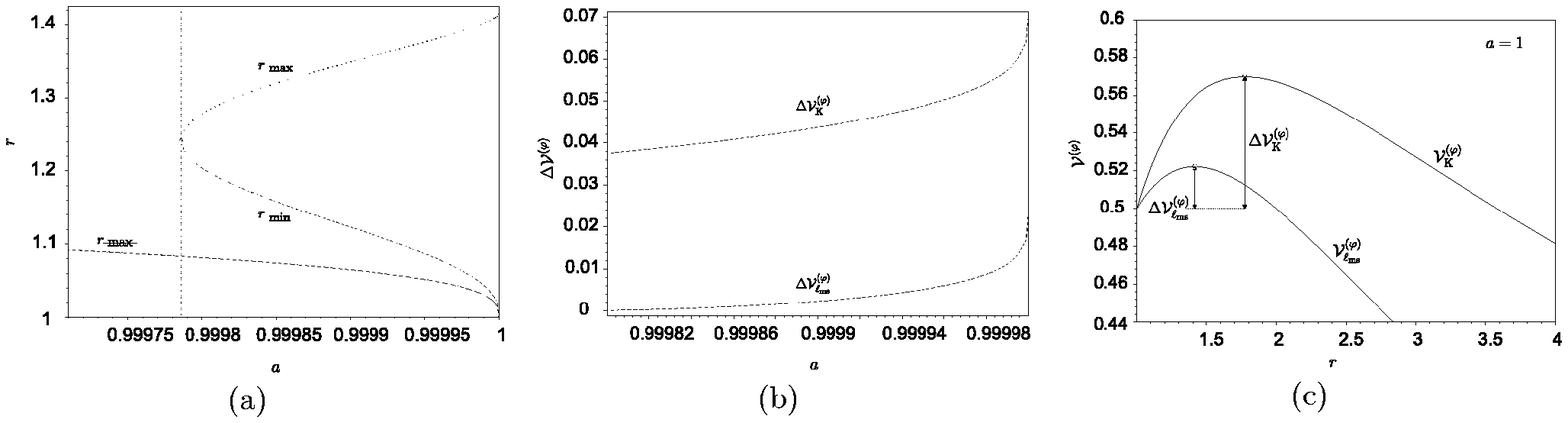}
\caption{(a) Positions of local extrema of ${\cal V}^{(\phi)}_{\rm
    LNRF}$ (in B-L coordinates) for the constant specific angular momentum tori
    with 
    $\ell=\ell_{\rm ms}$ in dependence on the rotational parameter $a$ of the
    black hole. (b) 
   Velocity difference $\Delta {\cal V}^{(\phi)}={\cal V}^{(\phi)}_{\rm
    max}-{\cal V}^{(\phi)}_{\rm min}$ as a function of the rotational
    parameter $a$ of the 
    black hole for both the Keplerian disc and the constant specific angular
    momentum   
    (non-Keplerian) disc with $\ell=\ell_{\rm ms}$. (c) Orbital-velocity
    curves in the limiting case of the extreme black hole. At $r=1$, the
    Keplerian orbital velocity ${\cal V}^{(\phi)}_{\rm K}$ has a local
    minimum, whereas the orbital velocity ${\cal V}^{(\phi)}_{\ell_{\rm ms}}$
    of the constant specific angular momentum disc has an inflex point. In both
    cases, the 
    velocity difference $\Delta {\cal V}^{(\phi)}$ reaches its maximal values:
    $\Delta {\cal V}^{(\phi)}_{\rm K}\doteq 0.06986,\ \Delta {\cal
    V}^{(\phi)}_{\ell_{\rm ms}}\doteq 0.02241$.}
\label{f7}
\end{figure*}

For $a=a_{\rm c(tori)}\doteq 0.99979$, except the irrelevant local maximum
located always 
outside the disc, an inflex point of ${\cal V}^{(\phi)}(r;a,\ell)$
occurs at $r_{\rm inf}\doteq 1.24143$ for the disc with $\ell=\ell_{\rm
  ms}\doteq 2.0123$. With rotational parameter growing ($a>a_{\rm c(tori)}$),
the local 
maximum of ${\cal V}^{(\phi)}(r;a,\ell)$ is succesively shifted up to values
of $r\sim 1.4$, while the local minimum of ${\cal V}^{(\phi)}(r;a,\ell)$ is
shifted down to $r=1$ in the limit of $a=1$ (Fig. \ref{f7}a). Notice that the
function ${\cal V}^{(\phi)}(r;a,\ell)$ can possess two local maxima even for
$a>a_{\rm c(bh)}\doteq 0.99964$ but for irrelevant values of the parameter
$\ell$, as it 
does not enters the interval corresponding to the constant specific angular
momentum tori,   
$\ell\not\in (\ell_{\rm ms},\ell_{\rm mb})$. The loci of these
extreme points can be directly inferred from Fig.~\ref{f4}, where the regions
corresponding to the constant specific angular momentum tori are shaded.  

For some representative cases corresponding to the classification of Kerr
spacetimes given in Sec. \ref{s3}, behaviour of ${\cal V}^{(\phi)}(r;a,\ell)$
is illustrated in Fig.~\ref{f6}, which enables us to make some conclusions on
the quantitative properties of the orbital velocity and its gradient. For
comparison, profiles of the Keplerian velocity ${\cal V}^{(\phi)}_{\rm
  K}(r;a)$ are included. With $a$ growing in the region of $a\in (a_{\rm
  c(tori)},1)$, the difference $\Delta {\cal V}^{(\phi)}\equiv {\cal
  V}^{(\phi)}_{\rm max}-{\cal V}^{(\phi)}_{\rm min}$ grows as well as the
difference of radii, $\Delta r \equiv r_{\rm max}-r_{\rm min}$, where the
local extrema of ${\cal V}^{(\phi)}(r;a,\ell)$ occur, see Figs.~\ref{f7}a,
b. Recall that the 
innermost local maximum of ${\cal V}^{(\phi)}(r;a,\ell)$ must be located,
necessarilly, under the disc structure. The value of ${\cal
  V}^{(\phi)}(r=r_{\rm in};a,\ell)$ at the inner edge of the toroid (where
$\ell=\ell_{\rm K}(r_{\rm in};a)$) is located closer and closer to the local
minimum of 
${\cal V}^{(\phi)}(r;a,\ell)$ when $a \to 1$. For $a=1$, there is an inflex
point of ${\cal V}^{(\phi)}(r;a=1,\ell)$ at $r=1$ where the local minimum and
the ``forbidden'' local maximum of ${\cal V}^{(\phi)}(r;a,\ell)$ for $\ell=2$
coincide, Fig.~\ref{f7}c. Moreover, the velocity difference $\Delta {\cal
  V}^{(\phi)}$ is smaller but comparable in the tori in comparison with
Keplerian discs. We can see that for $a \to 1$, the velocity difference in  
the tori $\Delta {\cal V}^{(\phi)}_{\rm tori}\approx 0.02$, while
for the Keplerian discs it goes even up to $\Delta {\cal V}^{(\phi)}_{\rm
  K}\approx 0.07$, see Fig.~\ref{f7}c. These are really huge velocity
differences, being expressed in units of $c$. 
 
We can conclude that the {\em Aschenbach effect}, i.e., the change of sign of
gradient of the LNRF orbital velocity, in the case of constant specific angular
momentum tori occur for the discs orbiting  
Kerr black holes with the rotational parameter $a>a_{\rm c(tori)}$. In terms
of the redefined rotational parameter, $1-a$, its value of 
$1-a_{\rm c(tori)}\doteq 2.1\times 10^{-4}$ is more than one order lower than
the value $1-a_{\rm c(K)}\doteq 4.7\times 10^{-3}$ found by Aschenbach for
the changes of sign of the gradient of orbital velocity in Keplerian
discs. In constant specific angular momentum tori, the {\em Aschenbach effect}
is elucidated by topology changes of the von Zeipel surfaces. In addition to
one self-crossing von Zeipel surface existing for all values of the rotational
parameter $a$, for $a>a_{\rm c(tori)}$ the second self-crossing surface
together with toroidal surfaces occur. Toroidal von Zeipel surfaces exist
under the newly developing cusp, being centered around the circle
corresponding to the minimum of the equatorial LNRF velocity profile.

\begin{acknowledgments}
The authors were supported by the Czech GA\v{C}R grants 202/02/0735 and
205/03/H144. The main parts of the work were done at the Department of
Astrophysics of Chalmers University at G\"{o}teborg and at Nordita at
Copenhagen. The authors Z.S.,
P.S. and G.T. would like to express their gratitude to the staff of the
Chalmers University and Nordita for perfect hospitality.
\end{acknowledgments}


\end{document}